# Longitudinal Spin Seebeck Effect in Silver Strip on CoFe Film

Y. Sheng,[1,2] M. Y. Yang,[1] Y. Cao,[3] K. M. Cai,[1] G. N. Wei,[1] G. H. Yu,[3] B. Zhang,[1] X. Q. Ma,[2,a)] and K. Y. Wang,[1,b)]

[1]*State Key Laboratory of Superlattices and Microstructures, Institute of Semiconductors, Chinese Academy of Sciences, Beijing 100083, China*

[2]*Department of Physics, University of Science and Technology Beijing, Beijing 100083, China*

[3]*Departement of materials physics, University of Science and Technology Beijing, Beijing 100083, China*

We report the experimental observation of the spin Seebeck effect (SSE) in Ag/CoFe noble metal/magnetic metal bilayers with a longitudinal structure. Thermal voltages jointly generated by the anomalous Nernst effect (ANE) and the SSE were detected across the Ag/CoFe/Cu strip with a perpendicular thermal gradient. To effectively separate the SSE and the ANE part of the thermal voltages, we compared the experimental results between the Ag/CoFe/Cu strip and Cu/CoFe/Cu strip, where two samples processed with the heating power instead of the temperature difference through the thin CoFe film. The respective contributions of the ANE and SSE to thermal voltage were determined, and they have the ratio of 4:1. The spin current injected through CoFe/Ag interface is calculated to be 1.76 mA/W.

There are three essential technologies in developing spintronics: the generation, detection and manipulation of spin currents.[1] A pure spin current can be realized by spin Hall effect[2,3], spin pumping[4,5] and spin Seebeck effect[6]. Pure spin current is beneficial for spintronic operation with much reduced energy dissipation. The spin Seebeck effect (SSE), which includes the spin current generation and detection, has attracted a lot of attentions since its first discovery in permalloy in 2008.[7] This effect consists of the generation of a spin

---

a) Electronic mail: xqma@sas.ustb.edu.cn.
b) Electronic mail: kywang@semi.ac.cn.

current as a result of an applied thermal gradient in a ferromagnet (FM), and a voltage detected by a noble metal (NM)[3] or magnetic metal[8] which has a strong spin orbital coupling attached on FM by means of the inverse Spin Hall Effect (ISHE). It offers applications of waste energy recovery or spin current generation, and a new point of view to discover the interplay between heat and spin currents in materials.

Materials including magnetic metals,[1] semiconductors,[9] and insulators[10] have been used to investigate the SSE in two structures, the transverse (TSSE)[11,12,13] with a temperature gradient applied in the sample plane and the longitudinal SSE (LSSE)[14,15,16] with a temperature gradient out of the sample plane, respectively. The longitudinal configuration is a more simple structure for measuring the SSE, and is good for high density integration. But the LSSE configuration is mainly used to investigate insulators other than the ferromagnetic metals, which are widely employed in spintronic applications such as the tunnel junction[17], spin valve[18] and memory devices[19]. When the FM is magnetic metal, the voltage of the SSE and anomalous Nernst effect (ANE)[20] in longitudinal structures are indistinguishable. The $E_{SHE} \propto j_S \times \vec{\sigma}$ due to SSE and $E_{ANE} \propto \nabla T \times \vec{m}$ with $\vec{m}$ owning the same direction with $\vec{\sigma}$ are collinear, where $E_{SHE}$, $j_S$, $\vec{\sigma}$, $E_{ANE}$, $\nabla T$, $\vec{m}$ are the electric field from the SHE, the spin current across FM/NM interface, the spin polarization of FM, the electric field of ANE, the temperature gradient of FM, and the magnetization of FM. Considering the ANE comes from the FM itself, the general way to distinguish these two effects is comparing two same FM layers with one adjoining to a NM layer and the other not, under the same temperature gradient.[21] However, the temperature gradient of the FM layer for LSSE structure is hardly to be measured precisely, leading to the inaccuracy of SSE signal analysis from the experiment.

In this letter, we provide a scheme to separate the ANE and SSE in Ag/CoFe/Cu sample unambiguously. We use the ferromagnetic metal CoFe thin film as the spin injector, and normal metal Ag layer as the spin detector, and the ANE and SSE are entangled with each other. In addition, we choose Cu/CoFe/Cu structure for comparison, in which the Ag layer was replaced by a Cu layer. We chose Cu for its week spin Hall effect and leading no SSE signal, thus only ANE exists. The thermal signals of two samples were measured, and compared after normalized by their heating power. The heating power was used to represent the temperature gradient which couldn't be precisely measured directly, because the thermal power can transform into the temperature gradient according to sample's thermal conductivity and geometry. We finally separate these two effects, and obtain the value of ANE, SSE current per Watt. Using the obtained SSE results, the spin current injected into FM/NM interface was also obtained.

Two samples were prepared for the present experiments following the procedures below: Ag(3 nm)/CoFe(5 nm)/Cu(2 nm) (sample 1) and Cu(3 nm)/CoFe(5 nm)/Cu(2 nm) (sample 2) multilayers were deposited by magnetron sputtering on a thermally oxidized Si substrate (0.5 mm in thickness) and patterned into a rectangle structure with the dimension $1\times 5$ mm$^2$ by Ar-ion beam etching.

The SSE was measured using the so called longitudinal configuration [see Fig. 1(a)]. The two samples were placed between two Cu plates as the hot source and cold source. To reduce the thermal contact and relieve the non-uniform heating between the Cu blocks and the sample, thermal grease was used between the Cu plate and the sample. A piece of insulating cotton was placed on the top of the hot side to make sure that the heat flux mainly flowed through the sample, as shown in the insert of Fig. 1(c). Two Pt-100 thermometers were placed on the two Cu plates to monitor the temperature. The thermal voltage was measured through two probes attached to both ends of the sample using a Keithley 2182A digital nanovoltmeter. All measurements were performed under vacuum (smaller than $1\times 10^{-4}$ mbar) in a cover of aluminum, in order to minimize thermal conduction through air, convection and disturb of electro-magnetic induction from electromagnetic wave. It would take 300-400 s until the temperature difference between the hot and the cold end becomes constant, and the temperature of both cold and hot sides keep slightly increasing parallelly for longer time, as shown in Fig. 1(b).

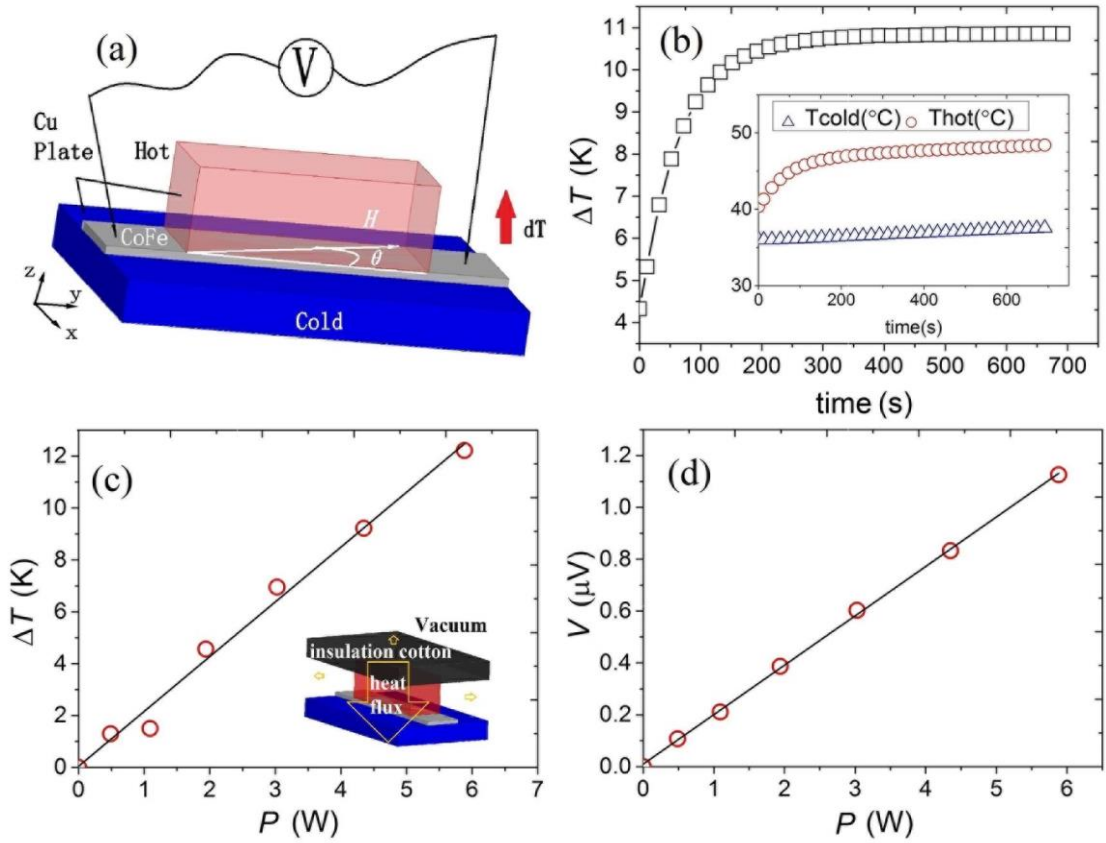

FIG. 1. (a) Structure of both sample 1 and sample 2, the red part means the hot side with dimension of $3\times 1 mm^2$, the blue part is the cold side and the grey part is Ag(Cu)/CoFe/Cu multilayers with dimension of $5\times 1\ mm^2$. The temperature gradient is out of plane, and the thermal voltage is measured along the length of FM layer. $\theta$ is the angle of in-plane magnetic field and the length direction. (b) The time-dependence of temperature difference $\Delta T$ between hot and cold sides of Cu plates; the insert shows the time-dependence of temperature on either hot or cold side, indicating that the temperature of both sides keep slightly increasing parallelly when $\Delta T$ is constant. As it is not necessary to take every measurement until $\Delta T$ comes to zero, $\Delta T$ doesn't begin with zero as the time does. Heating power $P$ dependence of $\Delta T$ (c) and $V$ (d) of sample 1 and their linear fitting curves are shown in the end.

Figure 1(c) shows the linear relation between the temperature difference of the two Cu plates ($\Delta T$) and the heating power ($P$). In vacuum condition, $P$ could be taken as the thermal power passing through the sample, as the heating power would transport mainly through the sample to the cold side, and only very small part will be dissipated by air and the insulating cotton. The relation between $P$ and $\Delta T_{FM}$ is $P=\kappa_{FM}S_{xy}\Delta T_{FM}/d$, where $\kappa_{FM}$ is the thermal conductivity of the FM layer, $d$ is the thickness of the FM layer. The accurate $\Delta T_{FM}$ can be obtained if $\kappa_{FM}$ is known. The $\kappa_{FM}$ of the two samples are equivalent because of the same magnetic material. Thus if the same heating power is applied to sample 1 and 2, the temperature gradient of the CoFe layer in these two samples will be the same. Figure 1(d) shows that the thermal voltages $V = [\Delta V(+H_S) - \Delta V(-H_S)]/2$ in sample 1 linearly increase with increasing the heating power, consisting with the linear $\Delta V - T$ relation in previous study[7,9,21,22], where $H_S$ is the saturated magnetic field applied in *x* direction. Besides the temperature gradient, the SSE is strongly dependent on the saturation magnetization ($M_S$) of the FM layer. In order to exclude the contribution from the change of $M_S$ during increasing of the heating power, we measured the magnetic hysteresis loops using Magneto optical Kerr effect when sweeping the magnetic field from -450 to 450 Oe in *x* direction at different temperature ranging from 25 to 78 ℃, as shown in the insert of Fig. 2. Figure 2 shows that the saturation magnetization is independent to the temperature in sample 1, indicating that $M_S$ is hardly influenced by the temperature in the range of measurement and a constant magnetization can be taken. Thus the $\kappa_{FM}$ is barely changed during the increase of heating power, and we can use the heating power instead of $\Delta T$ to normalize the thermal voltage when comparing the thermal voltages of the two samples.

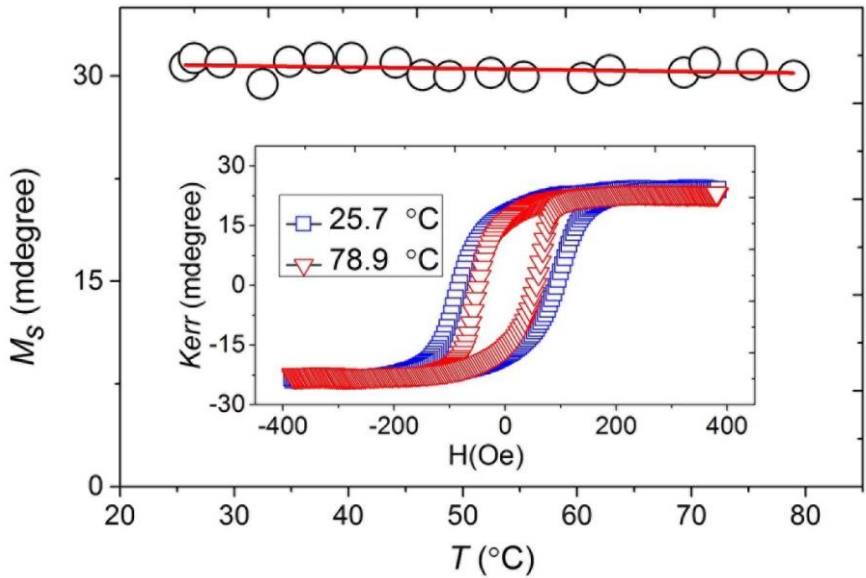

FIG. 2. Temperature dependence of the saturation magnetization of CoFe film. The insert shows the hysteresis loop at both the maximum (red inverted triangle) and minimum (blue square) temperature in measurement.

Figure 3(a) shows the thermal voltage hysteresis loops with the magnetic field applied along different in-plane orientations for sample 1, in which the classical Seebeck effect component was subtracted by moving the center of the loop to the origin point. Because the classical Seebeck effect coming from the non-uniform distribution of the temperature in *x-y* plane, was unchanged during sweeping the magnetic field. As a result of the existence of thermal gradient in *x-y* plane, the Planar Nernst effect (PNE) due to the scattering of electrons, which were driven by heat flux in the sample plane, would contribute to the thermal voltage. To analyze the PNE, the angular dependence of $S_m$ was shown in Fig. 3(b), where $S_m = V_m / P_m R_m$, in which $m = S1$ and $m = S2$ describe Sample 1 and 2 respectively. $S_m$ shares the same $\theta$ dependence with the thermal voltage. $V_{PNE}$ can be characterized by a $\sin(2\theta)$ dependence, while $V_{ANE}$ and $V_{SSE}$ are both $\sin(\theta)$ dependence. It is therefore possible to separate the $V_{PNE}$ and the other two terms by quantitatively analyzing the $V - \theta$ relation. In order to exclude the contribution of the magnetization change during the rotation of the magnetic field, the angular dependence of $V$ was measured at magnetic field well above the coercive field. The $S_m - \theta$ curve is obtained by averaging the $\Delta V$ values when the sample is fully magnetized. The angle dependence of $S_m$ for both the sample 1 and 2 are well fit to sine function, thus no PHE signal is detectable.

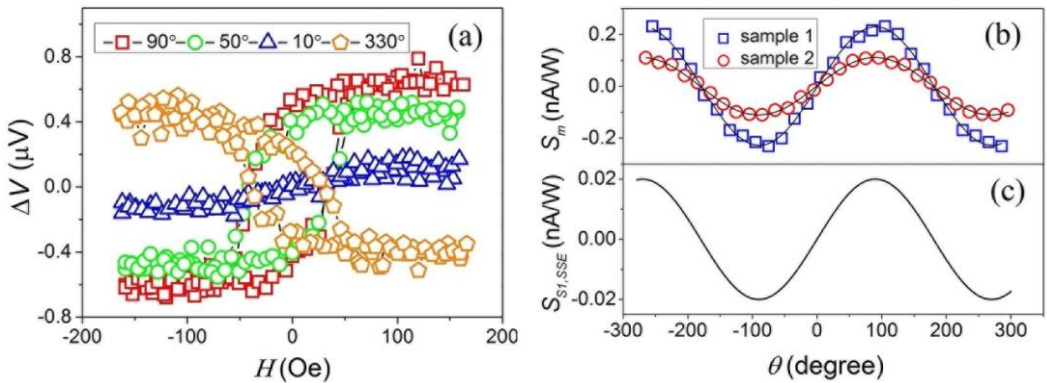

FIG. 3. (a) Hysteresis of thermal voltages $\Delta V$ across Sample 1 at fixed heating power with magnetic field applied along in-plane different orientations with angles: 90° (red square), 50° (green circle), 10° (blue triangle) and 330° (orange hexagon). (b) Angular dependence of $S_m$ in Sample 1 (blue square) and 2 (red circle), and solid lines are result

calculated by $B = A\sin(\theta)$. (c) Angular dependence of $S_{S1,SSE}$ in sample 1.

We proceed to separate the contribution of the ANE and SSE to the thermal voltage. Considering the expression from the electron transport theory: $J_m^i = \sigma_m^{ij} E^j - \alpha_m^{ik} \nabla_k T$, where $J_m^i$ stands for the electron current, E is the electric field, $\nabla_k T$ is the applied thermal gradient in *k*-direction, and the coefficients $\sigma_m^{ij}$ and $\alpha_m^{ik}$ are the elements of the conductivity and heating power tensor, respectively. We consider the current in *y* direction, which is an open circuit. The sources of the current can be divided into two groups: one is thermal gradient, and the other is thermal-induced electric field. The conductive current, resulting from the electric field due to the thermal effects, which force electrons to move directionally along the sample and accumulate on the edge of the sample, can be calculated by the voltage measured along *y* direction. The total currents of both sources are zero. We obtain the following expressions under our experimental condition:

$$J_{S1,con}^y = \frac{V_{S1}}{R_{S1}^y S_{xz}} = J_{S1,ANE}^y + J_{S1,SSE}^y, \tag{1}$$

$$J_{S2,con}^y = \frac{V_{S2}}{R_{S2}^y S_{xz}} = J_{S2,ANE}^y, \tag{2}$$

where $J_{S1,con}^y$ and $J_{S2,con}^y$ stands for the conductive current of sample 1 and 2 in *y* direction, $R_{S1}^y$ and $R_{S2}^y$ are the resistance of the heating area of sample 1 and 2, $J_{S1,con}^y$ and $J_{S2,con}^y$ are the conductive current of sample 1 and 2. $J_{S1,ANE}^y$ and $J_{S2,ANE}^y$ should be equal when the FM film of the two samples has the same temperature gradient. Assuming there is not any spin Hall effect in Cu, there is only ANE in sample 2, while both ANE and SSE exist in sample 1. Subtracting the results of sample 2 from sample 1, the $J_{S1,SSE}^y$ can be obtained accordingly.

To compare the thermal-induced current in two samples under the same heating power, we defined the $S_m$ mentioned above. This factor represents the value of the thermal induced current per Watt. Figure 3(b) shows the angular dependence of $S_m$ for both samples. The best fitting of the both curves follow sine relation, which give the amplitude $S_{S1} = 0.025$ nA/W and $S_{S2} = 0.005$ nA/W. The obtained $S_{S1}$ is much larger than that of $S_{S2}$, where the bottom metal layer silver in sample 1 has stronger spin orbital coupling than copper in sample 2. Thus the extra part of $S_m$ in sample 1 is attributed to the SSE, which is shown in Fig. 3(c). The maximum value of the SSE ($S_{SSE} = 0.02$ nA/W) can be achieved with the saturation magnetization direction along *x* axis. We can obtain the ratio of these two effects to be $V_{ANE}$ : $V_{SSE}$ =1 : 4, indicating that the SSE component is 3 times larger than the ANE. Considering the resistance value of 9.08 KΩ of sample 1, the voltage of SSE is 0.18 µV/W.

From the obtained $S_{SSE}$, the ability of the spin current generation at the Ag/CoFe interface in sample 1 can be calculated based on the expression[23] as follows :

$$j_S = \frac{t_{NM}\sigma_{NM} + t_{FM}\sigma_{FM}}{\theta_{SH}\lambda_{SD}\tanh(\frac{t_{NM}}{2\lambda_{SD}})} E_{ISHE} \tag{3}$$

where $t_{NM}$ ($t_{FM}$), $\sigma_{NM}$ ($\sigma_{FM}$), $\lambda_{SD}$, $\theta_{SH}$ and $E_{ISHE}$ is the thickness of the NM (FM) layer, electron conductivity of NM (FM) layer, the spin diffusion length, the spin hall angle and the electric field due to ISHE of the NM layer. As the $\lambda_{SD}$ of silver is 700 nm, which is much larger than $t_{NM}$ (3 nm), the $\tanh(t_{NM}/2\lambda_{SD})$ could be approximated to $t_{NM}/2\lambda_{SD}$, and the term of $\lambda_{SD}$ can be eliminated, which means spin current decreasing in the Ag strip could be ignored. So equation (3) can be simplified to $j_S = j_{S1,SSE}/\theta_{SH}$, and the $\theta_{SH}$ of silver is 0.0068,[24] and we can calculate the thermal-induced spin current per *Watt* in FM layer long *z* direction to be $I_S = j_S \cdot S_{xy} = 1.76$ mA/W, where $S_{xy}$ is the area spin current across.

In conclusion, we have observed the spin Seebeck effect in Ag/CoFe bilayers with longitudinal structure at room temperature. Normalizing the thermal voltage by heating power, thus we can directly compare the thermal voltages between the Ag/CoFe/Cu and Cu/CoFe/Cu samples. The contribution of SSE and ANE in Ag/CoFe/Cu sample is obtained to be $S_{SSE} = 0.02$ nA/W and $S_{ANE} = 0.005$ nA/W, respectively. The spin current injected into the silver layer is $I_S = 1.76$ mA/W, indicating the heating power of 1 Watt could generate a spin current of 1.76 mA in this structure. The method used in this paper is not only effective to separate the ANE and SSE in FM(metal)/NM structure, but also convenient to obtain the ability of spin current generation.

This work was supported by "973 program" No. 2014CB643903, NSFC Grant Nos. 61225021, 11174272 and 11474272.

[1] K. Uchida, T. Ota, K. Harii, K. Ando, H. Nakayama, and E. Saitoh, J. Appl. Phys. **107**, 09A951 (2010).

[2] S.O. Valenzuela and M. Tinkham, Nature **442**, 176 (2006).

[3] E. Saitoh, M. Ueda, H. Miyajima, and G. Tatara, Appl. Phys. Lett. **88**, 1 (2006).

[4] K. Ando, Y. Kajiwara, K. Sasage, K. Uchida, and E. Saitoh, IEEE Trans. Magn. **46**, 3694 (2010).

[5] H. Nakayama, K. Ando, K. Harii, Y. Fujikawa, Y. Kajiwara, T. Yoshino, and E. Saitoh, J. Phys. Conf. Ser. **266**, 4 (2011).

[6] J.-C. Le Breton, S. Sharma, H. Saito, S. Yuasa, and R. Jansen, Nature **475**, 82 (2011).

[7] K. Uchida, S. Takahashi, K. Harii, J. Ieda, W. Koshibae, K. Ando, S. Maekawa, and E. Saitoh, Nature **455**, 778 (2008).

[8] B.F. Miao, S.Y. Huang, D. Qu, and C.L. Chien, Phys. Rev. Lett. **111**, 1 (2013).

[9] C.M. Jaworski, J. Yang, S. Mack, D.D. Awschalom, J.P. Heremans, and R.C. Myers, Nat. Mater. **9**, 898 (2010).

[10] K. Uchida, J. Xiao, H. Adachi, J. Ohe, S. Takahashi, J. Ieda, T. Ota, Y. Kajiwara, H. Umezawa, H. Kawai, G.E.W. Bauer, S. Maekawa, and E. Saitoh, Nat. Mater. **9**, 894 (2010).

[11] M. Schmid, S. Srichandan, D. Meier, T. Kuschel, J.-M. Schmalhorst, M. Vogel, G. Reiss, C. Strunk, and C.H. Back, Phys. Rev. Lett. **111**, 187201 (2013).

[12] M. Schmid, S. Srichandan, D. Meier, T. Kuschel, M. Vogel, G. Reiss, C. Strunk, and C.H. Back, 2 (2013).

[13] S.H. Wang, L.K. Zou, J.W. Cai, B.G. Shen, and J.R. Sun, Phys. Rev. B - Condens. Matter Mater. Phys. **88**, 1 (2013).

[14] T. Kikkawa, K. Uchida, S. Daimon, Y. Shiomi, H. Adachi, Z. Qiu, D. Hou, X.-F. Jin, S. Maekawa, and E. Saitoh, Phys. Rev. B **88**, 214403 (2013).

[15] K. Uchida, T. Nonaka, T. Kikkawa, Y. Kajiwara, and E. Saitoh, Phys. Rev. B - Condens. Matter Mater. Phys. **87**, 1 (2013).

[16] T. Kikkawa, K. Uchida, Y. Shiomi, Z. Qiu, D. Hou, D. Tian, H. Nakayama, X.F. Jin, and E. Saitoh, Phys. Rev. Lett. **110**, 1 (2013).

[17] S.S.P. Parkin, C. Kaiser, A. Panchula, P.M. Rice, B. Hughes, M. Samant, and S.-H. Yang, Nat. Mater. **3**, 862 (2004).

[18] Z.H. Xiong, D. Wu, Z.V. Vardeny, and J. Shi, Nature **427**, 821 (2004).

[19] S.S.P. Parkin, M. Hayashi, and L. Thomas, Science **320**, 190 (2008).

[20] T. Miyasato, N. Abe, T. Fujii, a. Asamitsu, S. Onoda, Y. Onose, N. Nagaosa, and Y. Tokura, Phys. Rev. Lett. **99**, 1 (2007).

[21] R. Ramos, T. Kikkawa, K. Uchida, H. Adachi, I. Lucas, M.H. Aguirre, P. Algarabel, L. Morellón, S. Maekawa, E. Saitoh, and M.R. Ibarra, Appl. Phys. Lett. **102**, 7 (2013).

[22] K.I. Uchida, H. Adachi, T. Ota, H. Nakayama, S. Maekawa, and E. Saitoh, Appl. Phys. Lett. **97**, (2010).

[23] E. Shikoh, K. Ando, K. Kubo, E. Saitoh, T. Shinjo, and M. Shiraishi, Phys. Rev. Lett. **110**, 1 (2013).

[24] H.L. Wang, C.H. Du, Y. Pu, R. Adur, P.C. Hammel, and F.Y. Yang, Phys. Rev. Lett. **112**, 1 (2014).